# Chiral phonons in polar LiNbO$_3$

Hiroki Ueda[1,*,†], Abhishek Nag[1,*,†,‡], Carl P. Romao[2,3], Mirian García-Fernández[4], Ke-Jin Zhou[4,X], and Urs Staub[1,*]

[1]*Center for Photon Science, Paul Scherrer Institute, Villigen, Switzerland.*
[2]*Department of Materials, ETH Zurich, Zurich, Switzerland.*
[3]*Department of Materials, Faculty of Nuclear Sciences and Physical Engineering, Czech Technical University in Prague, Czech Republic.*
[4]*Diamond Light Source, Didcot, UK.*

[*]Correspondence authors: hiroki.ueda@psi.ch, abhishek.nag@ph.iitr.ac.in and urs.staub@psi.ch
[†]Equally contributed to this work.
[‡]Present address: Indian Institute of Technology, Roorkee, India.
[X]Present address: University of Science and Technology of China, Anhui, China.

## Abstract

Quasiparticles describe collective excitations in many-body systems, and their symmetry classification is of fundamental importance because they govern physical processes on various timescales, e.g., excited states, transport phenomena, and phase transitions. Recent studies have revealed that quasiparticles can possess chirality and that this degree of freedom leads to various important phenomena. Among them, chiral phonons have recently attracted significant interest because of their intrinsic "magnetism", which non-trivially bridges the spin system and the lattice. Here, we directly prove the presence of chiral phonons in a prototypical polar crystal LiNbO$_3$. Our demonstration adds a polar crystal in the showcase of materials hosting chiral phonons and, furthermore, creates a substantial potential in chiral phononics because of its expected *in-situ* switchable phonon chirality and associated control of phonon angular momentum.

**Main text**

**Introduction**

A phonon is one of the archetypical quasiparticles by which collective atomic motions can be described as the creation of a bosonic single particle. Even though a phonon is a concept used to describe lattice excitations, it has been demonstrated that a phonon can possess a mixed character with magnetism, opening the intriguing possibility of phononic control of magnetism. In general, such a hybridized mode can be classified as a dynamical multiferroic mode [1,2]. Examples include electromagnons [3,4] and phonons with angular momentum [5-8]. The latter is often referred to as a chiral phonon. The magnetic aspect of chiral phonons can be understood by the Barnett effect [9] at ultrafast timescales [10]. In equilibrium, the Barnett effect describes the induced magnetization in a spinning magnetically disordered medium. On ultrafast timescales, the magnetization is induced by the revolution of atoms, which could be either an intrinsic eigenmode [11,12] or coherent excitation of degenerate linear modes with a relative phase shift of $\pi/2$, e.g., triggered by a circularly polarized laser pulse in resonance with the degenerate phonon modes [2,10,13]. The emergent effective magnetic field from driving chiral phonons has significant potential to control magnetism at ultrafast timescales, e.g., as recently applied to magnetization switching [10] and coherent magnon excitation [14]. However, it is currently unclear why the induced field strength is gigantic compared to theoretical predictions [2,15], stimulating further development in understanding the "magnetism" of phonons [16-20] to form the basis of the new research field, chiral phononics.

Chiral phonons have attracted further interest from the opposite perspective, i.e., absorbing an angular momentum quantum from the magnetic system, which is not possible for conventional phonons. Recent experimental works have demonstrated that the angular momentum transfer occurs between spins and chiral phonons at ultrafast timescales, known as the ultrafast Einstein-de Haas effect [21,22], which is exactly the inverse effect of the ultrafast Barnett effect [10] and essential for the ultrafast demagnetization process. In addition, the creation of chiral phonons via magnon-phonon conversion highlights the critical role of phonon angular momentum in transport [23,24]. Thermal gradients create phonon angular momentum flow in a chiral crystal due to chiral acoustic phonons [25] and generate a spin current [26]. The substantial spin polarization of electrons propagating through a chiral crystal, known as chirality-induced spin selectivity and reaching an effective magnetic field in the order of 100 T [27], strongly indicates non-trivial coupling between electrons and chiral phonons. Besides, chiral phonons might mediate magnetic exchange interaction in heterostructures [28]. As such, the intrinsic "magnetism" of chiral phonons is spontaneously responsible for spin-lattice coupling and leads to various non-trivial phenomena and functionalities in materials.

Three different types of phonons have been referred to as chiral phonons in the community: (1) a rotational mode at the Γ point, such as the one reported in $SrTiO_3$ [2] and $CeF_3$ [13], (2) a rotational mode propagating in the rotation plane, such as the mode at the high symmetry points *K* and *K'* in transition metal dichalcogenides, e.g., $WSe_2$ [8], and (3) a rotational mode that propagates perpendicular to the rotation plane, such as the one observed in α-quartz [12], α-HgS [11], and tellurium [29]. All of them possess phonon angular momentum. However, from the symmetry point of view, e.g., applying the definition of chirality in dynamical objects [30], one finds that only the last type fulfills the symmetry requirements of being a dynamical chiral object, whereas the first two types are actually not chiral. Note that phonons of the second type are sometimes called two-dimensional chiral or cycloidal phonons. We, therefore, refer to only the last type as chiral phonons hereafter.

So far, chiral phonons have only been experimentally reported in chiral crystals, but this is not a general requirement [31]. For simplicity, let us first consider degenerate phonon modes in a crystal with preserved time-reversal symmetry as an example. Chiral phonons with opposite handedness can be expressed by phonon angular momentum **J** and phonon momentum **q**, which is either **J** // **q** (right-handed) or –**J** // **q** (left-handed). Thus, their inner product is a good quantity to describe phonon chirality [32]. Note that **J** is a time-odd axial vector while **q** is a time-odd polar vector. Their degeneracy at the Γ point is due to the time-reversal symmetry. Generally, at an arbitrary **q**, the time-reversal symmetry constrains a right-handed mode (–**J**, –**q**) to be equivalent in energy to a right-handed mode (**J**, **q**). The space-inversion operation connects a right-handed mode (–**J**, –**q**) to a left-handed mode (–**J**, **q**). Therefore, in centrosymmetric crystals, the presence of time-reversal symmetry results in the degeneracy of the two chiral phonon modes (**J**, **q**) and (–**J**, **q**) at all the momentum points, precluding chiral phonons or a chiral phonon band splitting. However, the degeneracy is lifted in non-centrosymmetric crystals, i.e., the absence of space-inversion symmetry (see Fig. 1**a**). Hence, this symmetry consideration imposes that space-inversion symmetry breaking is the crucial ingredient for the presence of chiral phonons and leads to a question: can one detect chiral phonons in a crystal that is nonchiral but breaks the space-inversion symmetry, e.g., a polar crystal? Theory predicted the presence of phonon angular momentum in ferroelectric $BaTiO_3$ [33]. However, $BaTiO_3$ forms ferroelastic domains in addition to ferroelectric domains, making the verification more challenging.

Here, we demonstrate the presence of chiral phonons in the prototypical polar crystal LiNbO$_3$ within a single ferroelectric domain by using resonant inelastic X-ray scattering (RIXS) with circular polarization, a sophisticated technique for probing chiral phonons [12]. Angular momentum transfer between a circularly polarized X-ray photon and a chiral phonon needs to fulfill the selection rules of phonon excitation in the RIXS process, which results in a circular polarization contrast on a chiral phonon excitation peak. LiNbO$_3$ possesses a polar corundum structure with the space group $R3c$ below the Curie temperature 1483 K [34], as shown in Figs. 1**b** and 1**c** (in hexagonal setting). The off-centering of Li$^+$ and Nb$^{5+}$ from the O$^{2-}$ triangle and octahedra, respectively, creates spontaneous polarization along $c$. Note that $R3c$ is a polar space group breaking the space-inversion symmetry but is not a chiral space group as it possesses $c$ glide symmetries. Its Brillouin zone viewed along $c^*$ is shown in Fig. 1**d**, together with the symmetry elements of $R3c$ and some momentum points where we collected RIXS spectra, which include $\mathbf{q}_1 = (0.1, -0.2, 1)$, $\mathbf{q}_2 = (0, -0.175, 1)$, and $\mathbf{q}_3 = (-0.1, -0.1, 1)$.

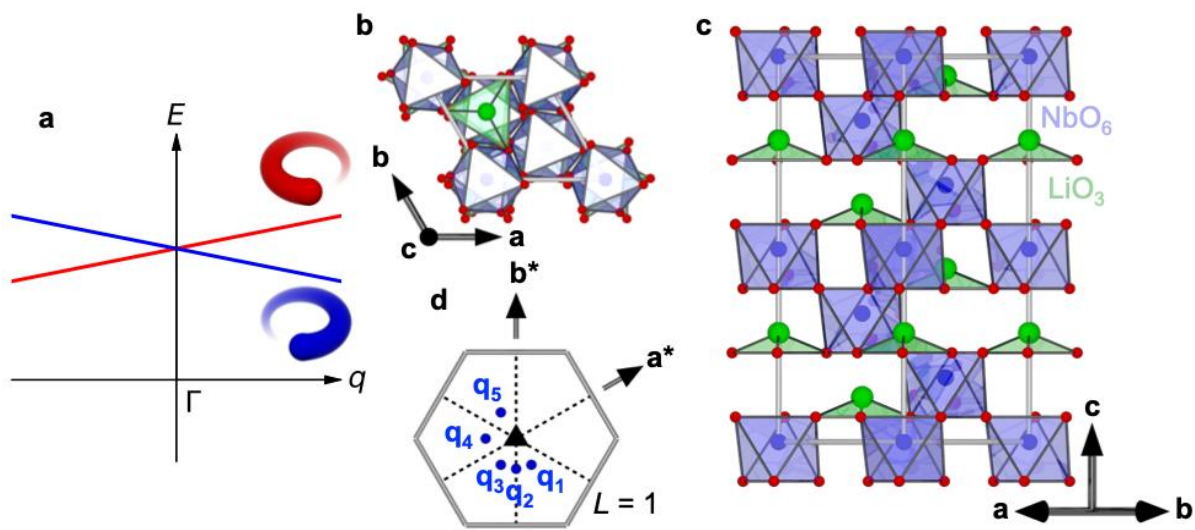

**Fig. 1 | Symmetry requirement of chiral phonons, and crystal structure and Brillouin zone of LiNbO$_3$. a**, Schematic drawing of chiral phonon dispersion with broken space-inversion symmetry and preserved time-reversal symmetry. Red and blue lines represent phonon dispersion with opposite phonon angular momenta. **b,c**, Crystal structure of LiNbO$_3$ in the hexagonal setting viewed along **b**, [001] and **c**, [110]. **d**, Brillouin zone of LiNbO$_3$ viewed along $c^*$ with momentum points where RIXS spectra have been collected. Here, $\mathbf{q}_1 = (0.1, -0.2, 1)$, $\mathbf{q}_2 = (0, -0.175, 1)$, $\mathbf{q}_3 = (-0.1, -0.1, 1)$, $\mathbf{q}_4 = (-0.2, 0.1, 1)$, and $\mathbf{q}_5 = (-0.1, 0.2, 1)$.

**Results and discussion**

Resonant X-ray scattering is described by a second-rank tensor and is sensitive to electric monopole (charge), magnetic dipole (spin), and electric quadrupole (orbital asphericity) for the dominant X-ray scattering process, i.e., electric dipole-electric dipole (E1-E1) transitions [35]. Figure 2a shows an X-ray absorption spectrum (XAS) around the O $K$ edge. At the O $K$ edge, a $1s$ core electron is excited into a $2p$ valence shell. Spin contributions are absent in diamagnetic LiNbO$_3$. In this case, circular dichroism (CD) in RIXS originates from the excitation of O $2p$ electric quadrupoles, as CD requires finite intensities in the polarization rotation channel in the X-ray scattering process, which is absent for isotropic charge scattering. As found in the RIXS energy map shown in Figs. S1b and S1e in Supplementary Information, phonon resonances are substantial for incident X-ray energies of ~530.85 eV and ~535.25 eV. The latter photon energy (represented by an arrow in Fig. 2a) is chosen for the RIXS measurements, as we do not find a clear CD-RIXS signal at the lower photon energy (compare Figs. S1c and S1f). The highest phonon energy in LiNbO$_3$ is ~110 meV [36], and all features above that energy are due to higher harmonics of phonon excitations.

According to Neumann's principle, chiral phonons in LiNbO$_3$ must respect the symmetry of $R3c$. Since a **J** component along a glide plane flips the sign by a glide operation, chiral phonons are not allowed at **q**$_2$ in the glide plane (i.e., $\mathbf{J}_{//\mathbf{q}} = \mathbf{0}$) but only those with a **J** component normal to the $c$ glide plane ($\mathbf{J}_{\perp\mathbf{q}} \neq \mathbf{0}$, two-dimensional chiral or cycloidal phonons). In contrast, since **q**$_1$ and **q**$_3$ are not in a $c$ glide plane but are connected by the glide symmetry, chiral phonons can exist ($\mathbf{J}_{//\mathbf{q}} \neq \mathbf{0}$) and, if they exist, must reverse their handedness between the two momentum points. In addition, the three-fold rotational symmetry along $c$ must be fulfilled. Figure 1d illustrates these relevant symmetry elements in reciprocal space, and Figs. S2a-S2c represent arrow plots of **J** corresponding to specific phonon modes obtained from density functional theory (DFT) calculations (described in Method).

Figures 2b-2d represent RIXS spectra taken at **q**$_1$, **q**$_2$, and **q**$_3$ with circular X-ray polarization (see Methods for details). There are three representative energy loss points in the spectra where we can find CD signals: ~25 meV ($E_1$), ~77 meV ($E_2$), and ~106 meV ($E_3$). CD signals are less substantial, e.g., ~7% for $E_2$ at **q**$_1$, compared to the one observed in α-quartz, ~17% [12]. In comparison to chiral α-quartz, LiNbO$_3$ has additional symmetry elements ($c$ glide planes) that constrain the appearance of nondegenerate chiral phonons, resulting in a generally reduced chiral phonon band splitting.

The RIXS CD is absent at **q**$_2$ and is roughly reversed between **q**$_1$ and **q**$_3$, as expected from the symmetry analysis, except for $E_3$. As details described in the Supplementary Information, the CD signal at $E_3$ seems significantly affected by the X-ray birefringence effect [37,38] because the modes at $E_3$ are almost pure linear translational and have large mode effective charge (see Fig. S6), resulting in a substantially linear dichroic RIXS amplitude (see Fig. S8). For the case here, the three-fold symmetry of the lattice and the two-fold symmetry of oscillatory electromagnetic waves make the birefringence six-fold symmetric. This results in identical CD between $q_1$ and $q_3$, in contrast to the three-fold rotational symmetry of chiral phonons, which reverses CD between the two momentum points (see detailed discussion in Supplementary Information). Therefore, one expects complex tangential momentum dependence of the RIXS CD at $E_3$. In fact, the RIXS CD at $E_3$ fits well with two sinusoidal functions that follow either the three-fold (chiral phonons) or six-fold (birefringence) rotational symmetry (see Fig. 3f). Our main discussion hereafter focuses on the modes at $E_1$ and $E_2$ unless otherwise stated.

Since CD on a phonon peak originates from angular momentum transfer between a circularly polarized photon and a chiral phonon, in principle, it could also appear for cycloidal (i.e., two-dimensional chiral) phonons. Due to the symmetry requirements, only **J** perpendicular to the $c$ glide plane, i.e., cycloidal phonons, can be finite at $q_2$, as found in the plots of phonon chirality, defined as **J**·**q** (see along $q_2$ in Figs. 3a-3c), and three-dimensional arrow plots of **J** (see Figs. S2a-2c in Supplementary Information) in reciprocal space obtained from DFT calculations. The absence of CD signals at $q_2$ indicates that the CD in our RIXS measurements is not sensitive to cycloidal phonons. This is likely because **J** given by a cycloidal phonon is perpendicular to the angular momentum of an incident circularly polarized X-ray photon in our experimental geometry, inhibiting an angular momentum transfer between the photon and cycloidal phonon (see Fig. S9 and Supplementary Information for details). This argument about the geometry being insensitive to cycloidal phonons also applies to the other momentum points where we have collected RIXS, making RIXS CD sensitive solely to chiral phonons.

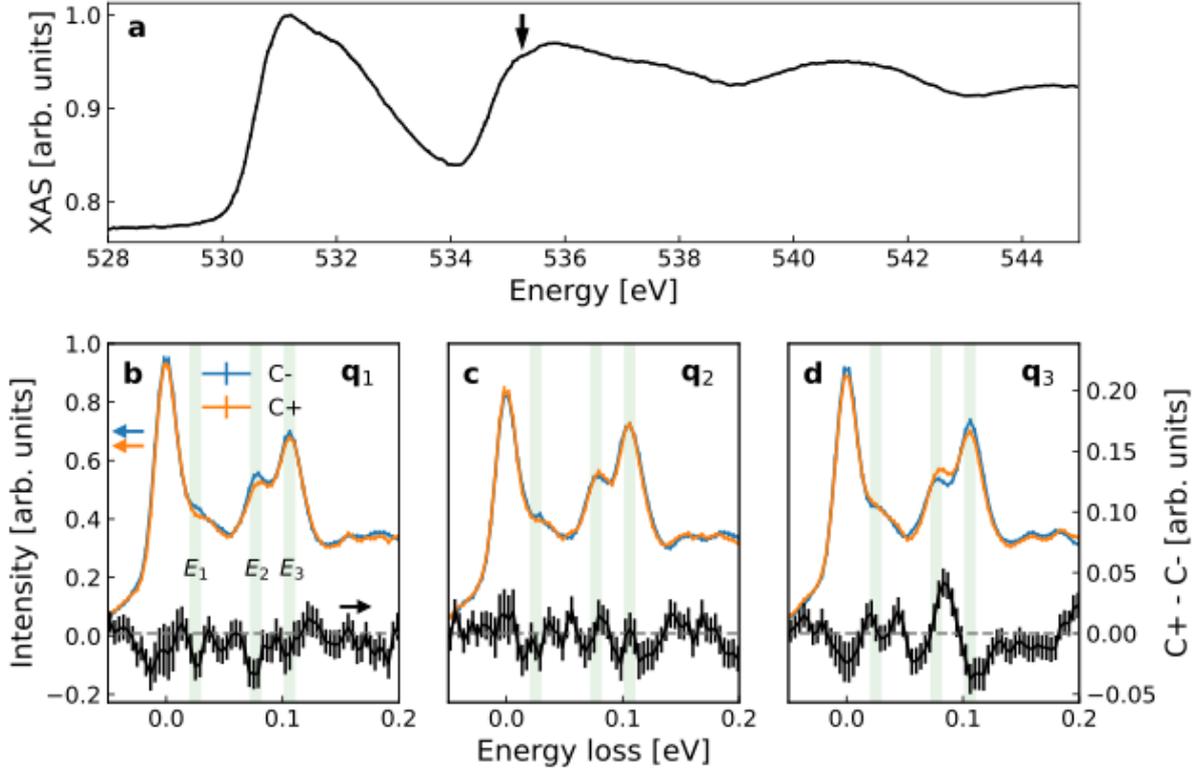

**Fig. 2 | XAS and RIXS. a**, XAS around the O $K$ edge. The arrow represents the photon energy for the RIXS measurements. **b-d**, RIXS with circular X-ray polarization at **b**, $q_1$, **c**, $q_2$, and **d**, $q_3$. The green bars highlight the representative energy loss points with finite CD, $E_1$, $E_2$, and $E_3$ (see text). The error bars in an RIXS spectrum are the standard deviation of individual scans.

The RIXS CD reversal between $q_1$ and $q_3$ at $E_1$ and $E_2$ (see Figs. 2**b** and 2**d**) is consistent with the plots of $\mathbf{J}\cdot\mathbf{q}$ (see Figs. 3**a** and 3**b**) and three-dimensional arrow plots of $\mathbf{J}$ (see Figs. S2**a** and S2**b** in Supplementary Information), indicating that the birefringence effect, which should be identical between the two momentum points, is less critical for these phonon modes. Polar plots of RIXS CD displayed in Figs. 3**d** and 3**e** confirm the three-fold rotational symmetry and its origin from the symmetry of the chiral phonons. Even though the birefringence with the six-fold rotational symmetry significantly affects RIXS CD at $E_3$, the finite contribution from the sinusoidal term with the three-fold rotational symmetry, as found in Fig. 3**f**, indicates that the mode at $E_3$ is also chiral. In fact, small but finite phonon chirality is predicted at $E_3$ by the DFT calculations, as shown in Fig. 3**c** and S2**c**.

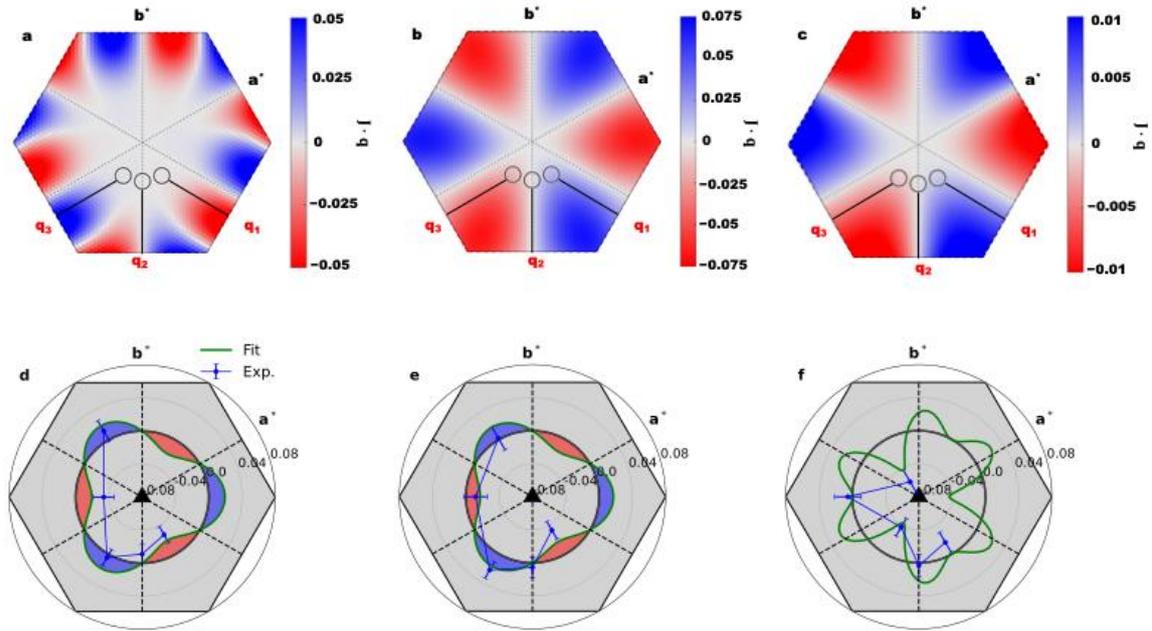

**Fig. 3 | Plots of phonon chirality and polar plots of circular dichroism in RIXS. a-c**, Surface plots of phonon chirality (**J·q**) centered at **a**, $E_1$, **b**, $E_2$, and **c**, $E_3$, shown for the plane corresponding to ($h\ k\ 1$) in hexagonal coordinates in reciprocal space. The plots consider contributions from each phonon weighted by a Gaussian centered at $E_1$, $E_2$, or $E_3$, with full width at half-maximum of 23 meV to account for the instrumental resolution. **d-f**, Polar plots of the RIXS CD for phonon modes at **d**, $E_1$, **e**, $E_2$, and **f**, $E_3$. The green curves are fits with either (**d** and **e**) a sinusoidal function with the three-fold rotational symmetry or (**f**) two sinusoidal functions that follows either the three-fold rotational symmetry (chiral phonons) or the six-fold rotational symmetry (birefringence). The error bars are propagated from RIXS spectra with opposite circular X-ray polarization with the error bars being the standard deviation of the individual scans.

Supplementary Videos 1-3 visualize the eigenmodes at the respective energies at $q_1$ viewed along $c^*$ by DFT calculations. All of them are circularly polarized, and their propagation involves the normal direction of the rotation plane, as also evident from the plots of **J·q** (see Figs. 3a-3c). Therefore, these phonons are clearly chiral. Note that a smaller projected **J·q** amplitude at $E_1$ than $E_2$ (compare Figs. 3a and 3b) despite clearer circularly polarized eigendisplacements for the mode at $E_1$ than $E_2$ (compare Supplementary Videos 1 and 2) is due to the convolution of a large number of bands with opposite chirality close in energy, as found in Fig. S3 in Supplementary Information. As in the case of α-quartz [12], the revolution of atoms distorts the O $2p$ electric quadrupoles by changing the Li-O-Nb bond angle, which gives rise to the sensitivity of RIXS to these phonon modes and its CD signals (see Fig. S6 showing mode effective charge of individual phonon bands). The induced effective magnetic moment from dynamical multiferroicity, considering only the circular motions of the Born effective charges, is in the order of sub-nuclear magneton, as shown in Fig. S4 in Supplementary Information, and is consistent with the α-quartz case [12] and previous DFT calculations [15].

## Conclusions and outlook

In conclusion, we have demonstrated the direct observation of chiral phonons in a polar crystal. As chiral phonons are a key ingredient of recently discovered exotic phenomena due to their phonon angular momentum [24-26], magnetism mediating the spin-lattice coupling [23,28,39], and non-trivial coupling with electrons [27], the demonstration of chiral phonons in a polar crystal is of significant value to the community as it adds new family members to the host materials of chiral phonons. In addition to these perspectives, our demonstration allows chiral phononics to go beyond the limitation of chiral crystals. Switching ferroelectric domains is straightforward, unlike chiral domains. Such domain inversion, in principle, allows us to switch spin-lattice coupling *in-situ* at general momentum points, opening unique opportunities to explore exotic phenomena based on controllable phonon chirality or phonon angular momentum. Based on the bi-stability of ferroelectric domains and the developed technology for local switching thereof [40], patterning chiral phononic devices in nanoscales will be possible. Synchronizing RIXS acquisition with the repetition of ferroelectric polarization reversal will directly map the phonon angular momentum as a function of ferroelectric atomic displacements. Of particular interest as a future perspective is the possible phonon angular momentum switching at ultrafast timescales due to the ultrafast reversal of ferroelectric polarization [41]. The controllable bistable states of phonon chirality or phonon angular momentum have significant potential to tailor emergent phenomena associated with chiral phonons at ultrafast timescales.

## Methods

**RIXS**

RIXS measurements were performed at Beamline I21 at the Diamond Light Source in the UK [42]. We tuned the photon energy to around the O *K* edge. The energy resolution was estimated as ~23 meV full width at half-maximum from the elastic peak of a carbon tape. A LiNbO$_3$ crystal with a single ferroelectric domain state and the largest face perpendicular to *c* in the hexagonal setting was commercially purchased. The manipulator at the beamline allows us to access different momentum points during the experiment. The error bars in an RIXS spectrum are the standard deviation of individual scans. The X-ray absorption spectrum was obtained by the partial fluorescence yield before the RIXS measurements.

**DFT**

Density functional perturbation theory calculations of the phonon frequencies and eigenvectors of LiNbO$_3$ were performed using the Abinit software package (v. 10) [43]. The calculations used the PBE GGA exchange–correlation functional [44] with the vdw-DFT-D3(BJ) dispersion correction [45]. The PAW method was used with a plane-wave basis set cutoff energy of 150 Ha within the PAW spheres and 30 Ha without. PAW basis sets were used as received from the Abinit library. A $5 \times 5 \times 5$ Monkhorst–Pack grid [46] was used to sample both **k**-points and **q**-points. The **k**-point grid spacing and plane-wave basis set cutoff energy were chosen following convergence studies, with the convergence criterion being 1% in pressure. Prior to the phonon calculation, the structure was relaxed to an internal pressure of −9 MPa, resulting in hexagonal lattice constants of $a = 5.13$ Å and $c = 13.9$ Å, in good agreement with experimental values ($a = 5.14$ Å and $c = 13.8$ Å) [47]. Calculations of the phonon circular polarization vectors and phonon magnetic moments were performed using a MATLAB script that is available as supplemental data.

## Acknowledgments

The resonant inelastic X-ray scattering experiment was performed at beamline I21 at the Diamond Light Source (proposal MM36210). C.P.R acknowledges support from the project FerrMion of the Ministry of Education, Youth and Sports, Czech Republic, co-funded by the European Union (CZ.02.01.01/00/22_008/0004591), the European Union and Horizon 2020 through grant no. 810451, and ETH Zurich. Computational resources were provided by the Swiss National Supercomputing Center (CSCS) under project ID s1128.

## Author contribution

H.U., A.N., and U.S. conceived and designed the project. H.U., A.N., M.G.F., K.-J.Z., and U.S. performed the RIXS experiment. H.U. and A.N. analyzed experimental data. C.P.R. performed density-functional theory calculations. H.U., A.N., C.P.R., and U.S. wrote the manuscript with contributions from all the authors.

## Competing interests

The authors declare no competing interests.

## Data availability

Experimental and model data are accessible from the PSI Public Data Repository [48].